\documentclass[letterpaper]{article} 
\usepackage[]{aaai25}  
\usepackage{times}  
\usepackage{helvet}  
\usepackage{courier}  
\usepackage[hyphens]{url}  
\usepackage{graphicx} 
\usepackage{natbib}  
\usepackage{caption} 
\usepackage{algorithm}
\usepackage{algorithmic}
\usepackage{newfloat}
\usepackage{listings}
\usepackage{adjustbox}
\usepackage{booktabs}
\usepackage{cite}
\usepackage[T1]{fontenc}
\usepackage{graphicx}
\usepackage[utf8]{inputenc}
\usepackage{inconsolata}
\usepackage{latexsym}
\usepackage{multirow}
\usepackage{microtype}
\usepackage{subcaption}
\usepackage{svg}
\usepackage{times}
\usepackage{xspace}

\usepackage{fancyvrb}
\usepackage[utf8]{inputenc}
\usepackage{xcolor} 
\usepackage[most]{tcolorbox} 
\usepackage{listings} 

\newtcolorbox{promptbox}[2][]{%
  enhanced,
  title=#2, 
  colback=gray!10,       
  colframe=black,        
  coltitle=white,        
  fonttitle=\bfseries,   
  boxrule=.5mm,          
  width=\linewidth,      
  arc=2mm
  sharp corners,         
  #1                     
}
\DeclareCaptionStyle{ruled}{labelfont=normalfont,labelsep=colon,strut=off} 
\floatstyle{ruled}
\newfloat{listing}{tb}{lst}{}
\floatname{listing}{Listing}
\newcommand{\rewritermethod}{{\sc Rewriter}\xspace}
\newcommand{\nl}{{\sc \textit{NL}}\xspace}
\newcommand{\db}{{\sc \textit{DB}}\xspace}
\newcommand{\sql}{{\sc \textit{SQL}}\xspace}
\newcommand{\nlsql}{{\sc \textit{NL2SQL}}\xspace}

\newcommand{\rewriter}{{\sc \texttt{Rewriter}}\xspace}
\newcommand{\reflector}{{\sc \texttt{Reflector}}\xspace}
\newcommand{\checker}{{\sc \texttt{Checker}}\xspace}

\newcommand{\memory}{{\sc \texttt{Memory}}\xspace}

\title{A Plug-and-Play Natural Language Rewriter for Natural Language to SQL}

\author {
    Peixian~Ma\textsuperscript{\rm 1},
    Boyan~Li\textsuperscript{\rm 1},
    Runzhi~Jiang\textsuperscript{\rm 1},
    Ju~Fan\textsuperscript{\rm 2},
    Nan~Tang\textsuperscript{\rm 1},
    Yuyu~Luo\textsuperscript{\rm 1}\thanks{Yuyu Luo is the corresponding author.}
}
\affiliations {
    \textsuperscript{\rm 1}The Hong Kong University of Science and Technology~(Guangzhou) \\
    \textsuperscript{\rm 2} Renmin University of China\\
    \{pma929, rjiang073@connect.hkust-gz.edu.cn,
    boyanli, nantang, yuyuluo\}@hkust-gz.edu.cn,
    fanj@ruc.edu.cn
}

\begin{document}

\maketitle

\begin{abstract}
Existing Natural Language to \sql~(\nlsql) solutions have made significant advancements, yet challenges persist in interpreting and translating \nl queries, primarily due to users' limited understanding of database schemas or memory biases toward specific table or column values. These challenges often result in incorrect \nlsql translations.
To address these issues, we propose \rewritermethod, a plug-and-play module designed to enhance \nlsql systems by automatically rewriting ambiguous or flawed \nl queries. By incorporating database knowledge and content (e.g., column values and foreign keys), \rewritermethod reduces errors caused by flawed \nl inputs and improves \sql generation accuracy. Our \rewritermethod treats \nlsql models as {\em black boxes}, ensuring compatibility with various \nlsql methods, including agent-based and rule-based \nlsql solutions.
\rewritermethod comprises three key components: \texttt{Checker}, \texttt{Reflector}, and \texttt{Rewriter}. The \texttt{Checker} identifies flawed \nl queries by assessing the correctness of the generated \sql, minimizing unnecessary rewriting and potential hallucinations. The \texttt{Reflector} analyzes and accumulates experience to identify issues in \nl queries, while the \texttt{Rewriter} revises the queries based on \texttt{Reflector}'s feedback.
Extensive experiments on the Spider and BIRD benchmarks demonstrate that \rewritermethod consistently enhances downstream models, achieving average improvements of 1.6\% and 2.0\% in execution accuracy, respectively.

\end{abstract}

\section{Introduction}

\begin{figure}[t!]
    \centering
    \begin{subfigure}[t]{0.44\textwidth}
        \includegraphics[width=\textwidth]{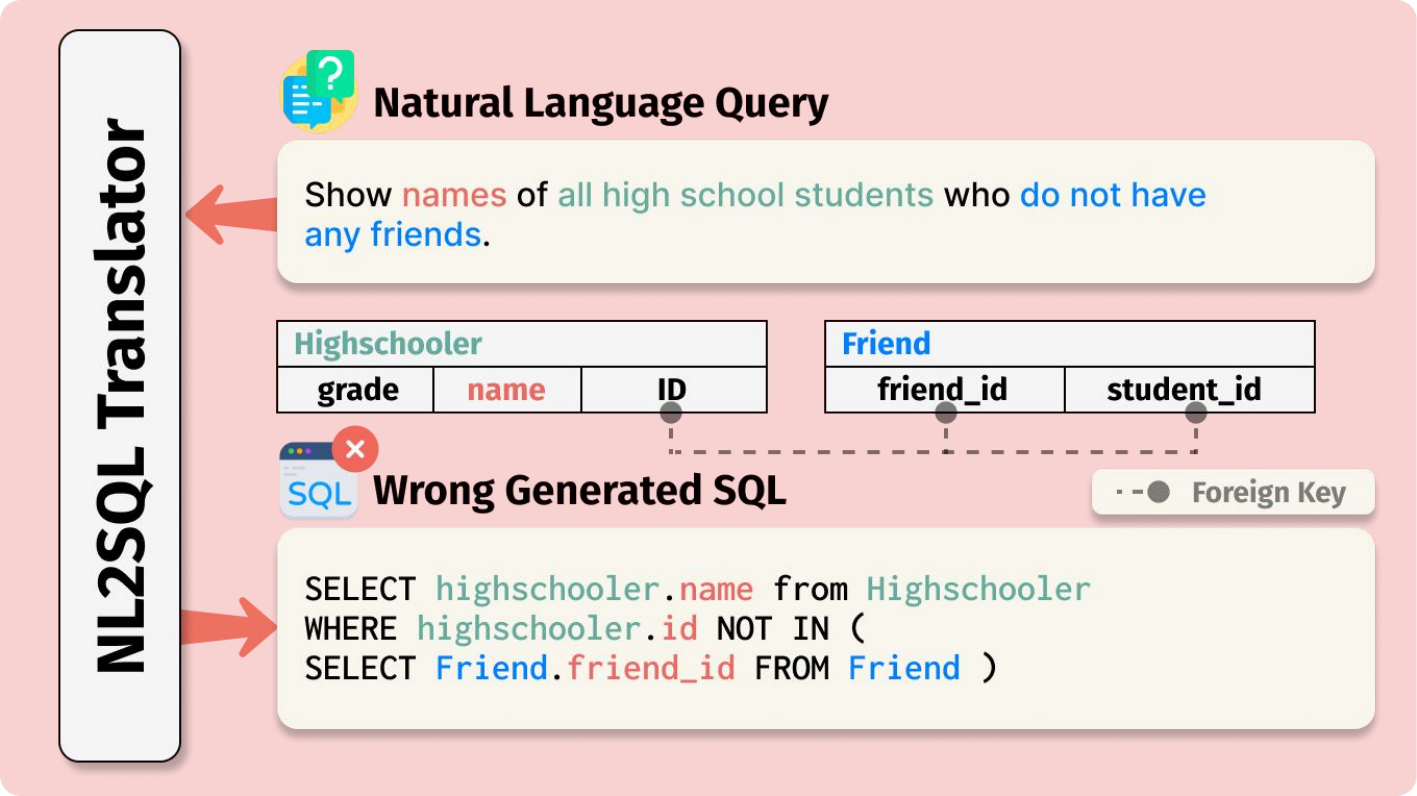}
        \caption{Traditional \nlsql system.}
        \label{fig:example_1}
    \end{subfigure}
    \begin{subfigure}[t]{0.44\textwidth}
    \includegraphics[width=\textwidth]{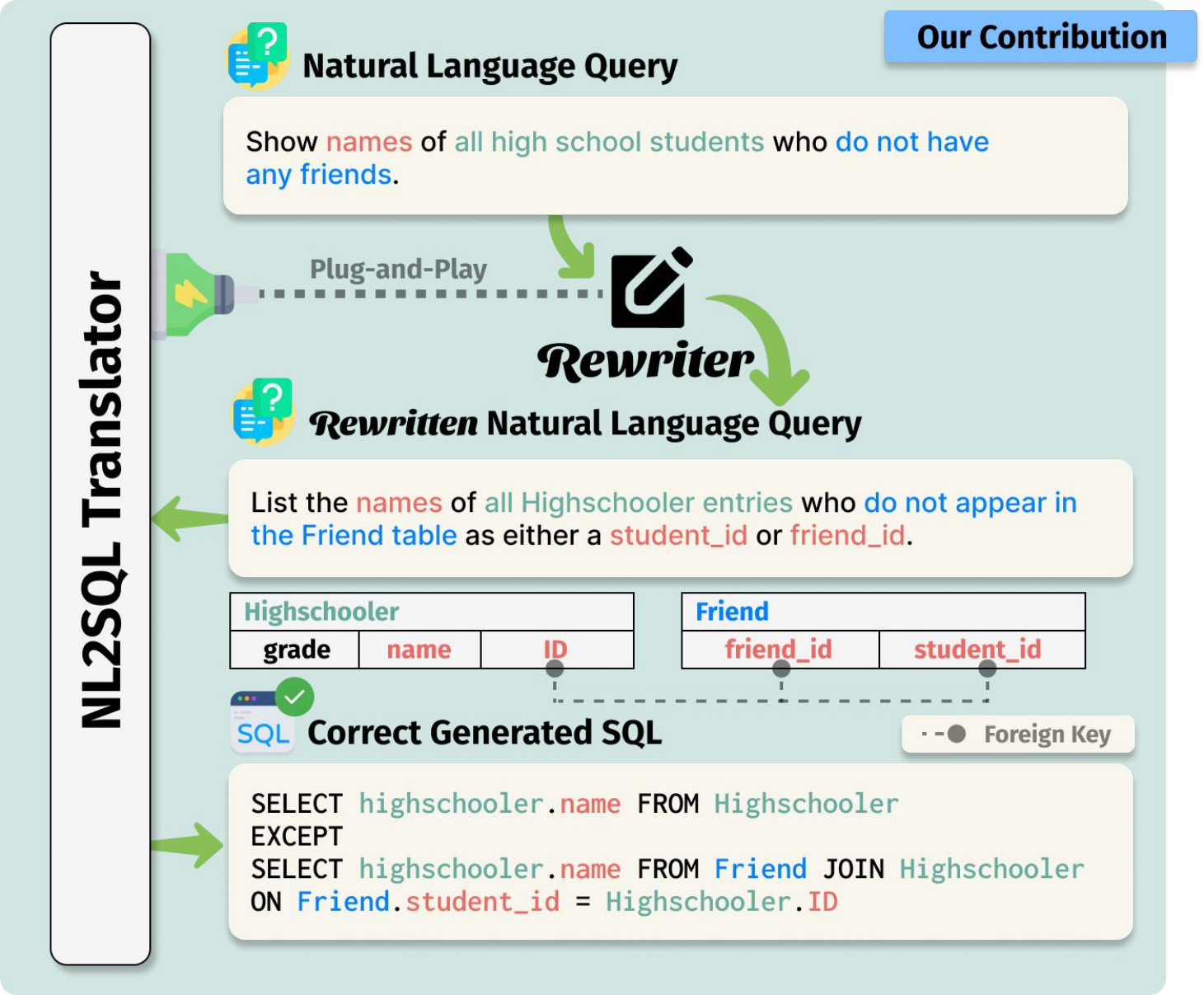}
        \caption{Deployment of a plug-and-play module to rewrite user \nl in above \nlsql system.}
        \label{fig:example_2}
    \end{subfigure}
    \caption{Demonstration of our work. The proposed plug-and-play \rewritermethod clearly indicates the relationships of foreign keys between tables in the rewritten \nl, thereby avoiding the generation of incorrect \sql statements due to the gap between unclear user intent and the \db.}
    \label{fig:example}
\end{figure}

\begin{figure*}[t!]
    \centering
    \includegraphics[width=0.95\textwidth]{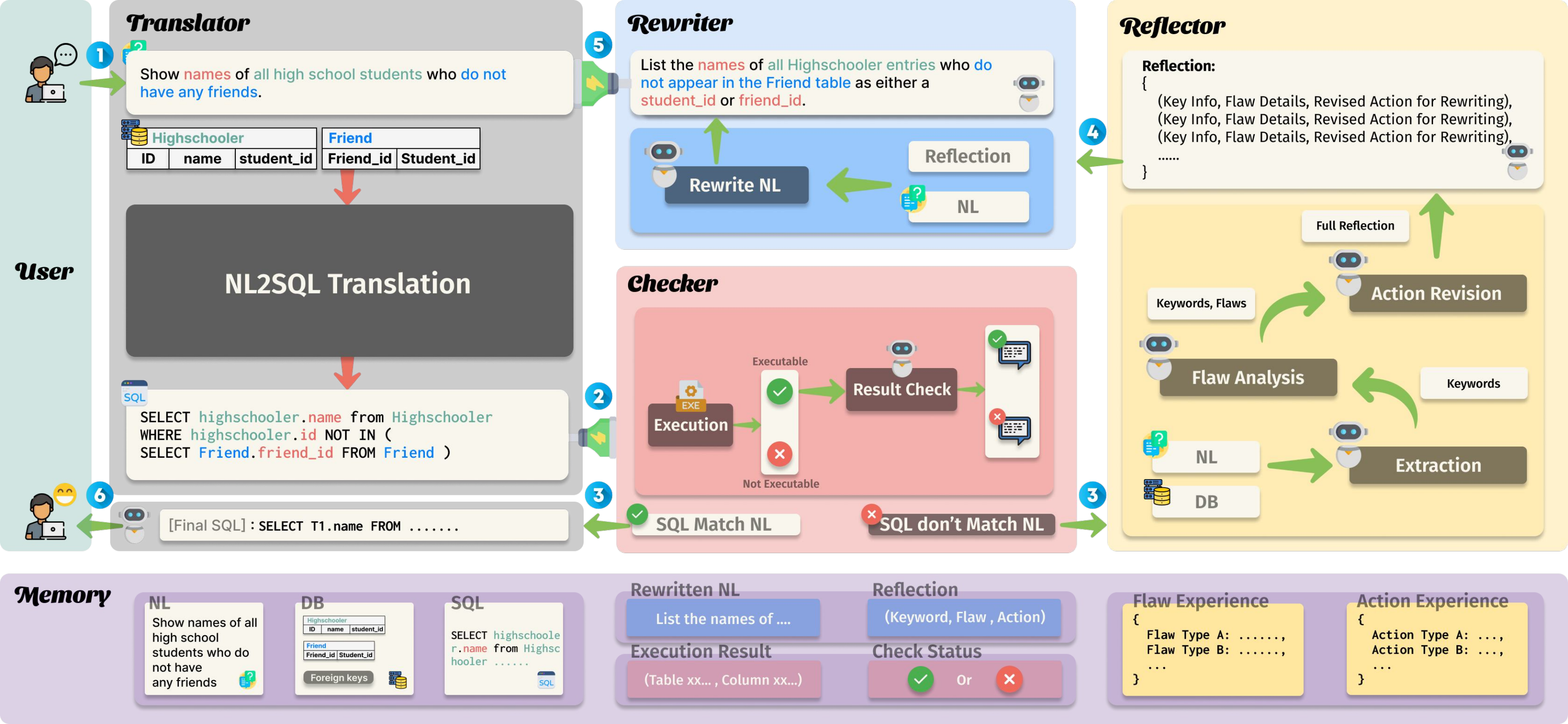}
    \caption{An overview of the proposed \rewritermethod framework, which comprises the following components: (i) \checker, which determines whether \nl matches the generated \sql; (ii) \reflector, which analyzes the flawed \nl and gives the rewriting reflection with the reference of \db; (iii) \rewriter, which rewrites the flawed \nl under the guidance of reflection. In addition, a task-specific \memory module provides information exchange and storage services for these agents.}
    \label{fig:model}
\end{figure*}

Natural Language to \sql~(\nlsql) allows users to convert natural language~(\nl) queries into structured \sql statements, simplifying database interaction without requiring \sql expertise~\cite{DBLP:journals/pvldb/LiLCLT24, DBLP:journals/corr/abs-2408-05109}, and supports a wide range of data science tasks~\cite{DBLP:conf/icde/LuoQ0018, DBLP:journals/debu/Luo00LZY20, DBLP:journals/tkde/LuoQCTLL22}.
Recent advancements in pre-trained and large language models (PLMs and LLMs) have significantly enhanced the semantic parsing capabilities of \nlsql systems, improving the integration of \nl and database content~(\db)~\cite{katsogiannis2023survey}. Recently, research efforts have primarily focused on optimizing various components of the \nlsql pipeline, including schema linking~\cite{resdsql}, database representation~\cite{chess}, encoding and decoding strategies~\cite{catsql, ratsql, codes}, and post-processing techniques~\cite{dinsql, macsql, c3sql}.

Despite these advancements, the quality of input \nl queries remains an underexplored challenge. In practice, users often lack sufficient knowledge of database schemas or struggle to express complex queries, particularly in multi-table scenarios. As depicted in Figure~\ref{fig:example_1}, this can result in ambiguous, incomplete, or flawed \nl queries, creating a significant gap between user intent and accurate \sql generation. 

To tackle this issue, we propose \rewritermethod, a plug-and-play framework designed to refine flawed \nl queries and enhance the overall \nlsql process. As shown in Figure~\ref{fig:example_2}, \rewritermethod leverages database knowledge (e.g., column values and foreign keys) to rewrite ambiguous or incomplete \nl queries, improving the performance of downstream \nlsql models.
Our \rewritermethod treats \nlsql models as {\em black boxes}, ensuring compatibility with diverse \nlsql solutions, including agent-based and LLM-based approaches.

However, our study presents several research challenges.

\textit{(C1) Lack of Training Data}.
Real-world datasets with flawed \nl queries and their corrected versions are scarce, making it difficult to directly train rewriting models.

\textit{(C2) Adaptability to Various \nlsql Solutions}. The varying capabilities of different \nlsql models pose a significant challenge. Rewriting strategies must be flexible enough to accommodate diverse systems, ranging from rule-based approaches to advanced LLM-based solutions, while ensuring consistency and robustness across varied user inputs and database schemas.

\textit{(C3) Minimizing Negative Impact}. Rewriting flawed \nl is a double-edged sword; while it can improve query clarity, it may also inadvertently degrade the performance of downstream \nlsql models. Therefore, it is crucial to carefully balance the benefits of rewriting with the risks of introducing errors or misinterpretations.

To address these challenges, we propose a multi-agent framework named \rewritermethod, which integrates three key components, i.e., \texttt{Reflector}, \texttt{Checker}, and \texttt{Rewriter}, supported by a shared \texttt{Memory} module that facilitates experience accumulation and exchange (see Figure~\ref{fig:model}).

To address the first challenge \textit{(C1)}, our \texttt{Reflector} employs a self-reflection mechanism that iteratively learns from its own trial-and-error process without relying on extensive labeled datasets. By analyzing \nl flaws and accumulating rewriting experiences, it extracts key details, identifies errors, and generates rewriting actions. It organizes this information into actionable reflections (e.g., keyword mismatches, missing information, or ambiguity) to guide precise revisions.

To address the second challenge \textit{(C2)}, the \texttt{Rewriter} focuses on targeted revisions informed by feedback from the \texttt{Reflector}, enhancing query clarity while maintaining the user’s original intent. In parallel, the \texttt{Checker} serves as a versatile evaluation module, ensuring alignment between \nl queries and generated \sql across a wide range of \nlsql systems, from rule-based to LLM-based approaches.

To minimize negative impact \textit{(C3)}, the \texttt{Checker} ensures that the rewriting process targets only flawed queries, avoiding unnecessary modifications to correct \nl. By executing the generated \sql on the database and comparing results to the user’s intent, it classifies queries as either valid (\texttt{SQL Match NL}) or flawed (\texttt{SQL Don’t Match NL}). This selective approach reduces the risk of over-rewriting, hallucinations, and errors.

\paragraph{Contributions.}
We make the following contributions.

\begin{itemize}
\item {\bf Plug-and-Play Framework.} We present \rewritermethod, a novel plug-and-play framework that identifies and rewrites ambiguous or flawed \nl queries in \nlsql systems. The framework integrates seamlessly with diverse \nlsql methodologies, including rule-based and LLM-based solutions.

\item \textbf{Self-Reflective Learning.} \rewritermethod incorporates a self-reflection mechanism to iteratively refine rewriting actions without relying on large-scale training data. Its conservative rewriting strategies ensure improved query clarity while minimizing potential negative impacts on downstream models.

\item \textbf{Comprehensive Evaluation.} Through extensive experiments on multiple benchmarks, we demonstrate that \rewritermethod enhances the performance of various \nlsql models. Our results highlight the framework’s adaptability and ability to learn effectively from experience.
\end{itemize}

\section{Methodology}
\subsection{Overview}
In this section, we propose \rewritermethod, a plug-and-play multi-agent framework designed to optimize user \nl in \nlsql task.
As illustrated in Figure \ref{fig:model}, \rewritermethod consists of the following LLM-based agents: the \checker for discriminating samples that \nl does not match \sql, the \reflector for analyzing the flawed \nl and developing rewrite actions, and the \rewriter for rewriting flawed \nl under the guidance of \reflector. Additionally, the \memory component stores key information throughout the rewriting process and facilitates communication among the agents.

\subsection{Checker}
\checker plugs the output port of the \nlsql model. The primary objective of it is to identify the samples that the generated \sql query does not match the user's \nl.
As demonstrated in Figure \ref{fig:model}, the workflow of \checker can be delineated into two distinct phases. 
In the first phase, \checker executes the generated \sql queries in the database and filters out those that have syntax errors or cannot be executed correctly. These samples will be marked as \texttt{NL DO NOT MATCH SQL}.
If the generated SQL query executes successfully, it will be included in the instructions for the second stage of the Checker agent. This will also comprise the related \nl statement, \db, and the corresponding query result. These rules focus not only on checking \sql syntax but also on identifying potential semantic errors, reference errors, and other issues among the \nl, \db, \sql, and query results.
In the second stage, the \checker agent assesses whether the query result aligns with the intent of the provided \nl based on the above information and professional verification rules. Samples that pass both stages of verification will be marked as \texttt{NL MATCH SQL} while those that do not will be labeled as \texttt{NL DO NOT MATCH SQL} and stored accordingly.

\subsection{Reflector} 

\begin{figure}[t!]
    \centering
    \includegraphics[width=0.8\columnwidth]{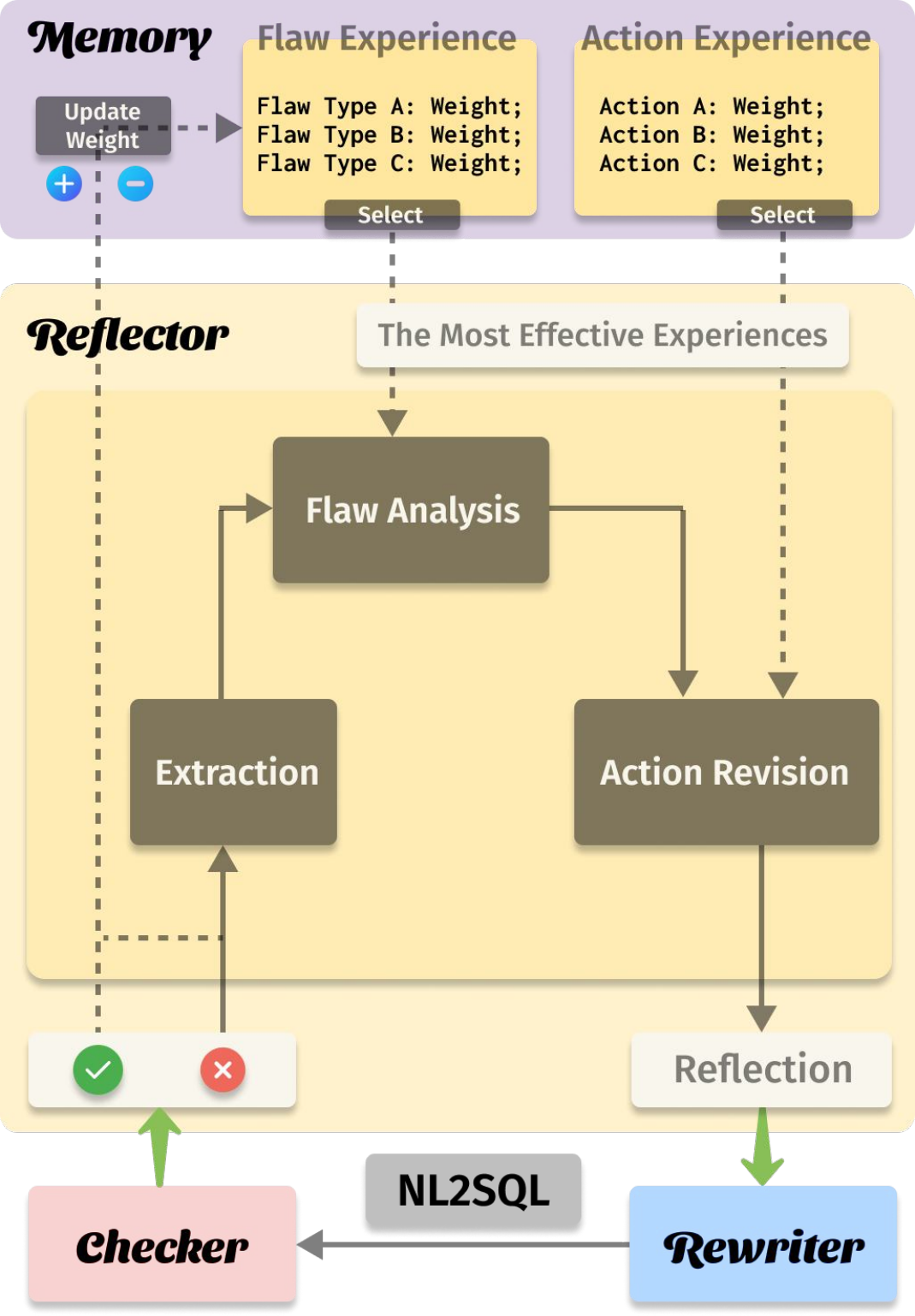}
    \caption{Demonstration of self-reflection mechanism. In the rewriting process, the experiences with the highest weights will be loaded into the rules.  Subsequently, the \reflector updates the weights of the experience, which is applied in the detailed reflection based on the checker's feedback on the new results.}
    \label{fig:reflection}
\end{figure}

\reflector works as the core of the \rewritermethod framework. In the process of reasoning,
it firstly extracts the key information (e.g. keywords or description of related tables and columns) in given \nl and compares the above information with \db.
Then, it verifies the key information with the provided \db~(e.g., Table names and column names, foreign keys) and then analyzes the information for any potential flaws or mismatch of user's intent~(e.g., Ambiguity or Missing information).
For example, \reflector examines the foreign keys in the database to determine if there is an intention to query multiple tables and columns.
Finally, specific rewriting suggestions for the above flaws are generated with reference to the above \db and action experience.
The final reflection output is collated as a triples text (\texttt{Keyword}, \texttt{Flaw}, \texttt{Action}), where the \texttt{Flaw} represents the flawed analysis of the key information and the \texttt{Action} represents the related rewriting suggestions.

Due to the lack of sufficient labeled training data, the \reflector is unable to summarize practical flaw experience or action experience through training or fine-tuning.    
Consequently, as shown in Figure \ref{fig:reflection}, we employ a self-reflection mechanism for updating the above experience.
In the initialization stage, the \reflector is instructed to generate various potential flaws of the user's \nl and related rewriting actions as the initialized experiences, which are condensed and not related to concrete database content or values.  
These experiences are stored in the \memory with associated weights corresponding to their validity and importance.
During the process of flaw analysis and action revision, a batch of experiences with the highest weight is selected as the most likely defect and the most effective rewriting action in the current \nl for the reference of \reflector.
Subsequently, the \reflector utilizes these experiences and \db to provide concrete flaw descriptions and revision suggestions pertaining to database values and concatenate them as the final reflection text.
Additionally, it marks the specific experiences used in the reflection.
Before the next round of reasoning and rewriting, the \reflector will adjust the weights of the above experience based on the reward signal, which is characterized by a sparse binary state (\texttt{NL MATCH SQL} or \texttt{NL DO NOT MATCH SQL}) provided by the \checker.
By continuously looping through the above process at reasoning, \reflector can accumulate some effective flaw experiences and action experiences, enabling it to adjust itself in time based on the distribution of flaw occurrences in different \nl., enhancing the robustness of rewriting \nl.

\subsection{Rewriter}
The task of \rewriter is to rewrite the specified flawed \nl and plug the rewritten \nl into the \nlsql translator.  
It retrieves the reflection text provided by the \reflector and the related \db from the \memory and performs the rewrite action guided by that information and specific rules.
In addition,  \rewriter also seeks to maintain consistency of narrative style between the rewritten \nl and the original \nl, which can avoid creating new hallucinations and misunderstandings for the \nlsql translation. The \rewriter will only attempt to modify the narrative style when the reflection explicitly states that a comprehensive modification of the expressive semantics is necessary.

\subsection{Memory}
The function of \memory is to store the data stream record of the above agents, such as \nl, \db, \sql, or details of flaws and actions summarized by \reflector.
\memory can continuously accumulate historical data from different \nlsql methods during the workflow of \rewritermethod, and support the agents to adapt to different \nl representation styles, thus avoiding dependence on a single rewrite strategy and improving the robustness of the rewritten \nl.

\section{Experiments}

\subsection{Settings}
\paragraph{Datasets} We evaluated the proposed \rewritermethod and related \nlsql translation models on two benchmarks, Spider~\cite{spider} and BIRD~\cite{bird}. 
Spider comprises 10,181 questions paired with 5,693 complex \sql queries from 200 databases and 138 domains.
BIRD~\cite{bird} comprises 12,751 \nlsql pairs encompassing 95 databases from 37 specialized domains. 

\begin{table}[t!]
    \centering
    \caption{\nlsql translation in Spider-dev set.}
    \label{table:spider}
    \begin{tabular}{p{4.8cm}ccc}
        \toprule
        \textbf{\nlsql Translation Methods} & \textbf{CP} & \textbf{EX} & \textbf{EM} \\ 
         \midrule
         C3~+~GPT-3.5-turbo                     & -     & 81.9          & 46.9 \\
         \hspace{1em} + \rewritermethod         & 55.8  & \textbf{82.4} & \textbf{47.0} \\
         DAIL-SQL~+~GPT-3.5-turbo               & -     & 76.3          & 60.5 \\
         \hspace{1em} + \rewritermethod         & 78.3  & \textbf{78.1} & \textbf{61.2} \\
         DAIL-SQL~+~GPT-4                       & -     & 83.1          & 70.0 \\
         \hspace{1em} + \rewritermethod         & 60.5  & \textbf{83.6} & \textbf{70.0} \\
         DTS-SQL + Deepseek-6.7B                & -     & 75.2          & 77.2 \\
         \hspace{1em} + \rewritermethod         & 66.0  & \textbf{77.5} & 47.9          \\
         NatSQL~+~T5-Base         & -     & 69.7          & 65.3 \\
         \hspace{1em} + \rewritermethod         & 81.3  & \textbf{72.2} & \textbf{68.5} \\
         NatSQL~+~T5-3B           & -     & 71.4          & 68.0 \\
         \hspace{1em} + \rewritermethod         & 82.2  & \textbf{75.7} & \textbf{71.9} \\
         RESDSQL + T5-3B                        & -     & 81.8          & 78.1 \\
         \hspace{1em} + \rewritermethod         & 57.0  & \textbf{82.2} & \textbf{78.6} \\
         \bottomrule
    \end{tabular}
\end{table}

\begin{table}[t!]
    \centering
     \caption{\nlsql translation results in BIRD-dev set.}
    \label{table:bird}
    \begin{tabular}{p{4.8cm}ccc}
        \toprule
        \textbf{\nlsql Translation Methods} & \textbf{CP} & \textbf{EX} & \textbf{VES} \\ 
         \midrule
         DAIL-SQL + GPT-4                       & -     & 54.3          & 56.1 \\
         \hspace{1em} + \rewritermethod         & 53.0  & \textbf{54.6} & 56.0 \\
         GPT-4                                  & -     & 46.4          & 49.8 \\
         \hspace{1em} + \rewritermethod         & 58.4  & \textbf{49.9} & \textbf{56.9} \\
         GPT-3.5-Turbo                          & -     & 36.6          & 43.8 \\
         \hspace{1em} + \rewritermethod         & 60.6  & \textbf{38.9} & \textbf{44.0} \\
         \bottomrule
    \end{tabular}
\end{table}

\begin{figure}[t!]
    \centering
    \includegraphics[width=0.9\columnwidth]{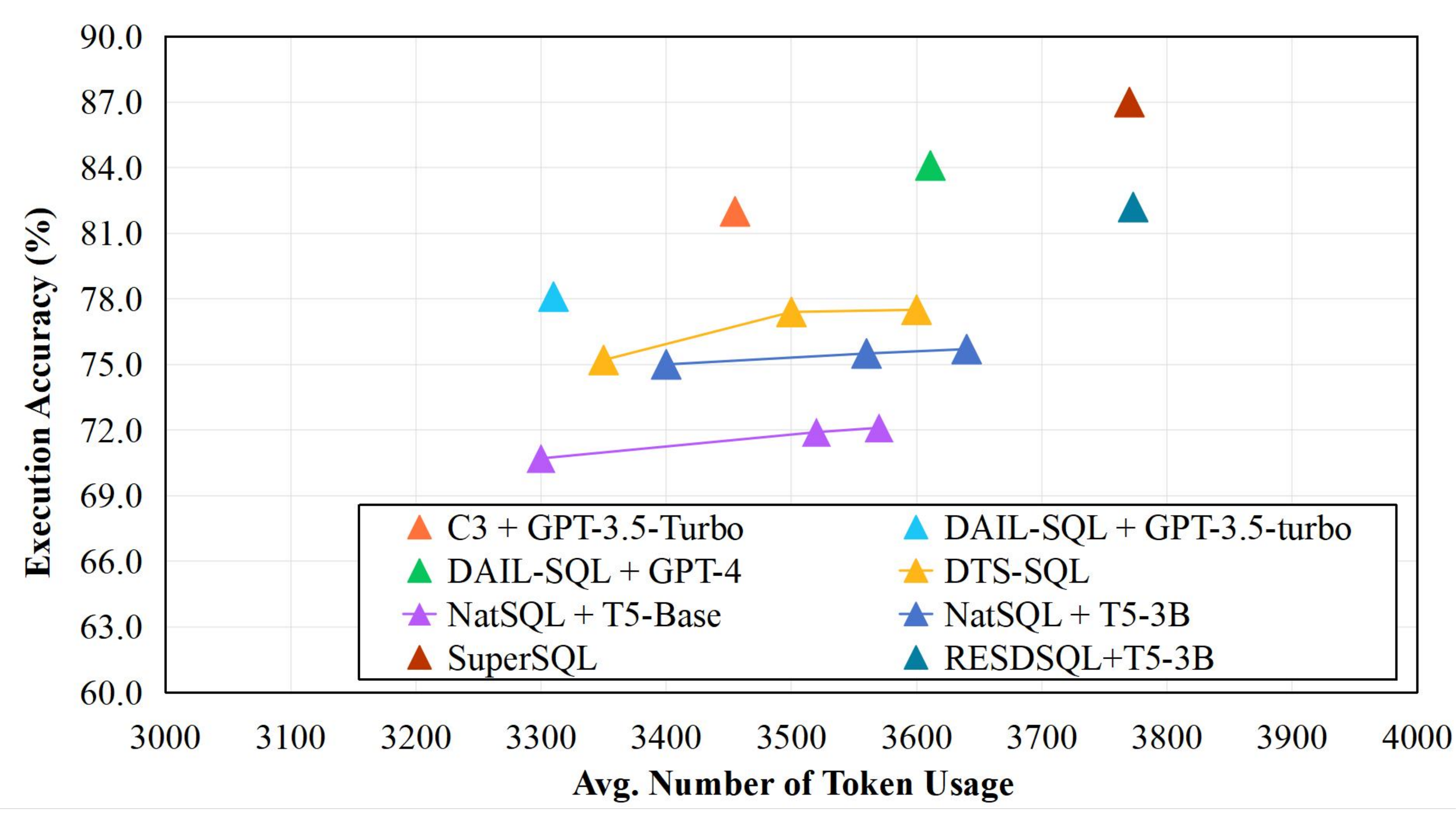}
    \caption{Execution accuracy and token efficiency of \rewritermethod in single-round and multi-round rewriting on Spider-dev set.}
    \label{fig:multi_round}
\end{figure}

\paragraph{Metrics} For fair comparisons, we follow the standard evaluation metric of each benchmark. 
For Spider, we use Execution Accuracy~(EX) and Exact Match Accuracy~(EM) as the metric. 
For BIRD, we utilize EX and Valid Efficiency Score~(VES) as the metric. 
EX is used to estimate the percentage of questions that predict the same result for the query and the basic gold query across all query requests.
EM focuses on measuring the exact match between model predictions and actual results. It treats each clause as a set and compares the prediction of each clause with the corresponding clause in the reference query.
VES is a metric that evaluates the efficacy of \sql queries generated by the \nlsql model. It specifically assesses the validity of the \sql queries by comparing their result sets with those of basic gold \sql queries. 
For the evaluation of the \checker, we utilized Check Precision~(CP) as the metric. CP is defined as the precision of the bad samples that \nl does not match the \sql.

\paragraph{Base Models} To compare the efficiency of different base models on agent performance, we select several open-source and closed-source LLM for comparison experiments, including GPT-4o, GPT-4~\cite{openaigpt4}, GLM-4~\cite{glm4} and LLaMA-3-8B~\cite{llama3}.

\paragraph{NL2SQL Translator Baselines} To comprehensively evaluate the performance of the \rewritermethod, we conduct experiments on both BIRD and Spider benchmark by plugging \rewritermethod on the following \nlsql translator baselines: 
C3~\cite{c3sql}, DTS-SQL~\cite{dtssql}, NatSQL~+~T5~\cite{comp}, RESDSQL~\cite{resdsql} are specific to Spider dataset; 
GPT-3.5 \& GPT-4~\cite{openaigpt4} are specific to BIRD dataset;
DAIL-SQL~\cite{dailsql} is specific to both datasets.

\paragraph{Environments} We conduct all experiments of \rewritermethod on a server with one NVIDIA A40 48GB GPU, one Intel(R) Xeon(R) Platinum 8358P CPU, 512 GB memory and Ubuntu 20.04.2 LTS operating system.

\subsection{Overall Performance}

\paragraph{SQL Generation Performance} Table \ref{table:spider} and Table \ref{table:bird}  illustrate the performance of the proposed \rewritermethod with various \nlsql baseline methods in the dev set of Spider and BIRD benchmark.
For most \sql generation results of the baselines, the \checker can achieve the CP of over 50\% in terms of the bad samples.      
The \sql queries generated by the rewritten \nl outperform those of the relevant baseline results in terms of execution accuracy, exact match accuracy, and valid efficiency score on both \nlsql benchmarks.  
It is noticeable that most \sql generation results with \rewritermethod rewriting part of flawed \nl are able to outperform the EX score of the baseline results by 1\%-3\%.
In addition, we also explore the performance of multi-round rewriting. As demonstrated in Figure \ref{fig:multi_round}, we observe that multiple rounds of rewriting partial flawed \nl can effectively improve the accuracy of \sql generation on some \nlsql methods.

\paragraph{Token Efficiency} For the agents in the \rewritermethod, we utilize different LLM as their base model.
In light of the fact that the LLM APIs incur charges based on the token count and that the inference time of the LLM is directly correlated to the token length, we aim to minimize the computational cost associated with rewriting while upholding the quality of the rewritten \nl. The total number of tokens is mainly influenced by the input prompt, including the representation of database content and the agent-specific rules. 
Our experiments evaluate the token efficiency of the \checker with different LLMs, which comprise close-source LLM and open-source LLM.
To compute the cost of the agent, we repeat the experiment 5 times and calculate the average token usage of its response.
Due to the lack of representative rewritten samples, all experiments are performed with zero-shot. 

\begin{figure*}[t!]
    \centering
    \begin{subfigure}[b]{0.48\textwidth}
        \includegraphics[width=\textwidth]{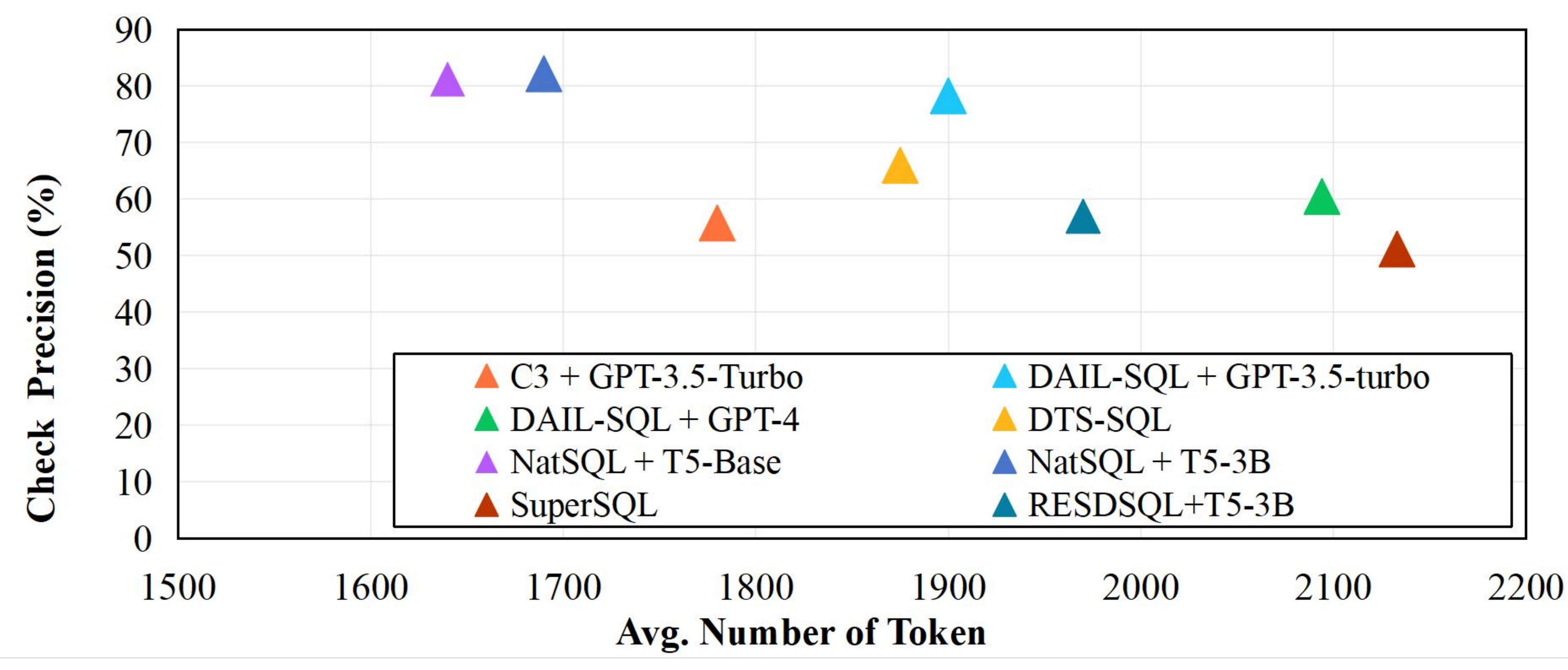}
        \caption{GPT-4o}
        \label{fig:sub1}
    \end{subfigure}
    \begin{subfigure}[b]{0.48\textwidth}
        \includegraphics[width=\textwidth]{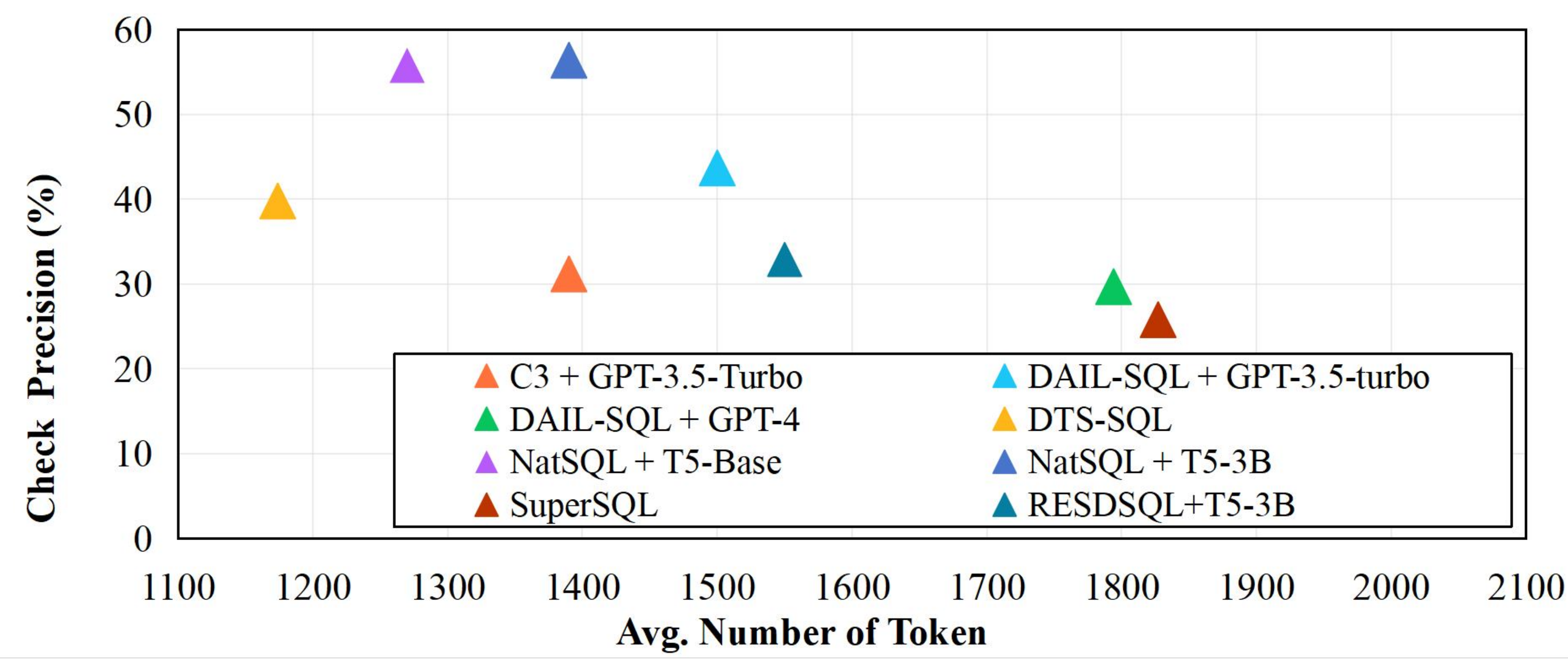}
        \caption{GPT-4}
        \label{fig:sub2}
    \end{subfigure}
    \begin{subfigure}[b]{0.48\textwidth}
        \includegraphics[width=\textwidth]{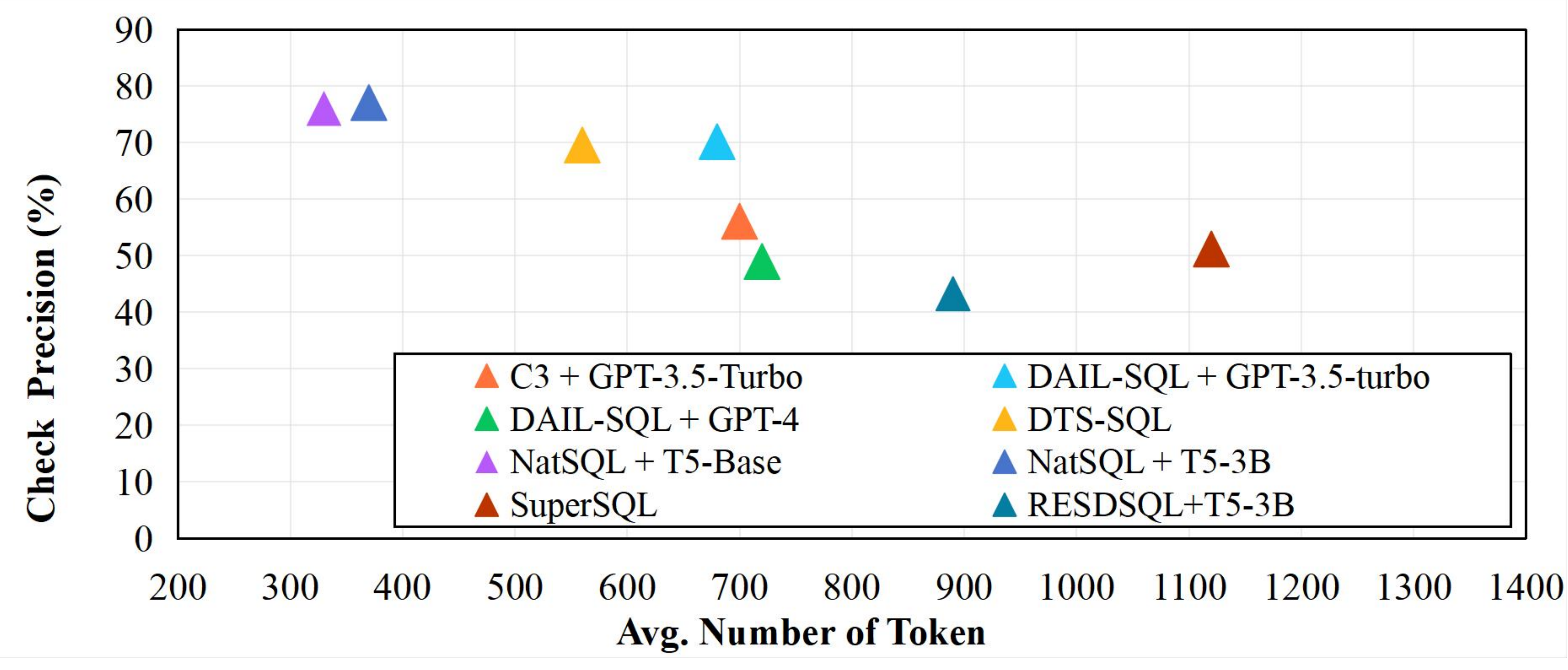}
        \caption{GLM-4}
        \label{fig:sub3}
    \end{subfigure}
    \begin{subfigure}[b]{0.48\textwidth}
        \includegraphics[width=\textwidth]{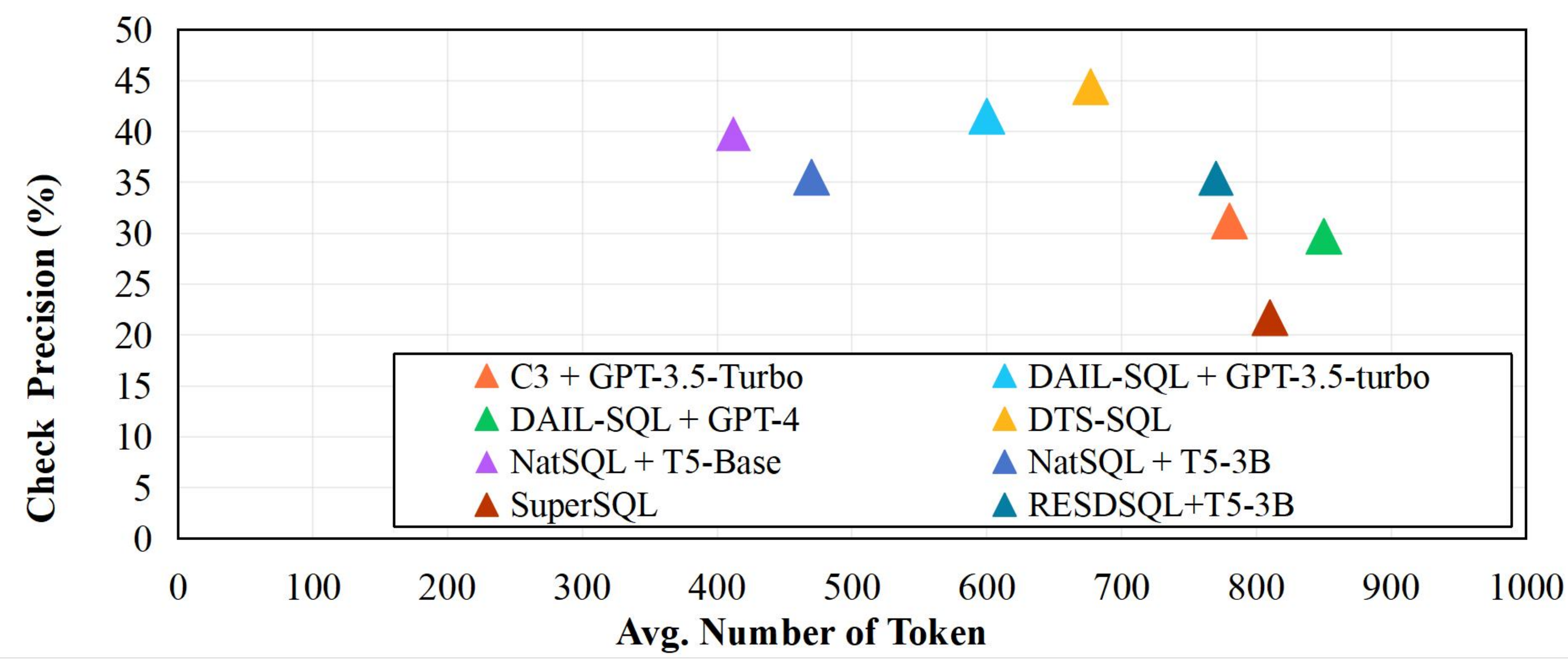}
        \caption{LLaMA-3-8B}
        \label{fig:sub4}
    \end{subfigure}
    \caption{Effectiveness of \checker.  
    Token efficiency vs. Precision on Spider-dev set.}
    \label{fig:token_efficiency}
\end{figure*}

With the above settings, Figure \ref{fig:token_efficiency} illustrates the performance and token efficiency of the \checker by employing different base models.
In the comparison of closed-source LLMS, GPT-4o achieves more than 50\% CP on the result of all \nlsql baselines, and the token cost is close to that of GPT-4 and GLM-4. Although open-source LLMs demonstrate the potential to minimize token cost, their performance lags behind that of closed-source LLMs. The deficiency in available training data and the absence of domain knowledge may contribute to this performance disparity.
In addition, Figure~\ref{fig:multi_round} also shows the impact of multi-round rewriting on token usage. Since the increasing length of rewritten \nl may cause the token increment in the whole workflow. However, due to the subsequent rounds only processing the bad samples, the overall token usage is relatively reduced.

\begin{table}[t!]
    \centering
    \caption{Ablation study on Spider-dev set. CK represents the \checker. RE represents the \reflector.}
    \label{table:ablation_agent}
        \begin{tabular}{lp{0.5cm}p{0.9cm}p{0.9cm}p{0.6cm}}
        \toprule
        \textbf{\nlsql Methods} & \textbf{Base} & \textbf{w/o CK} & \textbf{w/o RE}& \textbf{All} \\
        \midrule
        C3 + GPT-3.5-Turbo          & 81.9 & 78.8(↓) & 80.8(↓)  & \textbf{82.4} \\
        DAIL + GPT-3.5-Turbo        & 76.3 & 76.9(↑) & 77.7(↑)  & \textbf{78.1} \\
        DAIL + GPT-4                & 83.1 & 78.1(↓) & 81.8(↓)  & \textbf{83.6} \\
        DTS-SQL                     & 75.2 & 71.0(↓) & 74.4(↓)  & \textbf{77.5}\\
        NatSQL + T5-Base            & 69.4 & 70.2(↑) & 71.5(↑)  & \textbf{72.2} \\
        NatSQL + T5-3B              & 71.8 & 71.6(↓) & 72.9(↑)  & \textbf{75.7} \\
        RESDSQL                     & 81.8 & 77.3(↓) & 80.4(↓)  & \textbf{82.2} \\
        \bottomrule    
        \end{tabular}
\end{table}

\begin{table}[t!]
    \centering
    \caption{Ablation study on Spider-dev set. HC. represents the hand-craft experiences, IN. represents the self-initialized experiences, LE. represents the learning-updated experiences.}
    \label{table:ablation_reflector}
        \begin{tabular}{lp{0.5cm}p{1.1cm}p{1.1cm}p{0.6cm}}
        \toprule
        \textbf{\nlsql Methods} & \textbf{Base} & \textbf{HC.} & \textbf{IN.}& \textbf{LE.} \\
        \midrule
        DTS-SQL                 & 75.2 & 77.2 (↑) & 76.4 (↑) & \textbf{77.5}\\
        NatSQL + T5-Base        & 69.4 & 70.7 (↑) & 72.1 (↑) & \textbf{72.2}\\
        NatSQL + T5-3B          & 71.8 & 75.0 (↑) & 73.7 (↑) & \textbf{75.7}\\
        \bottomrule    
        \end{tabular}
\end{table}

\subsection{Ablation Study}
We perform several ablation studies for  \rewritermethod. 
\subsubsection{Ablation Study for Agents} 
Table \ref{table:ablation_agent} illustrates the results of the ablation study for the \rewritermethod, where w/o \checker represents the rewriting of all \nl in the dataset, and w/o \reflector represents instructing \rewriter to rewrite in the absence of analysis. 
The experimental results demonstrate that the \checker and \reflector within the \rewritermethod significantly contribute to the whole rewrite action.  
Removing any one of these agents can result in a decrease in the quality of rewritten \nl. 
In addition, Table \ref{table:spider} and Table \ref{table:ablation_agent} illustrate that rewriting all \nl may lead to new hallucinations, which negatively affect the \nlsql model. The \checker can effectively detect the wrong \sql, so as to ensure that the \rewritermethod can process the real flaw \nl as much as possible,  minimizing the new hallucinations caused by rewriting the correct \nl.

\subsubsection{Ablation Study for Self-reflection} 
Table \ref{table:ablation_reflector} presents the ablation experiment on \reflector. In this study, we examine the performance of \reflector in various settings, encompassing the utilization of hand-crafted experiences, self-initialized experiences, and learning-updated experiences. 
The hand-crafted experience refers to the flaws and rewriting action schemes summarized by NL2SQL experts;
The self-initialized experience refers to related experience generated at the initialization stage of \reflector;
The learning-updated experience refers to the experience that the \reflector summarizes and updates in the multi-round rewriting process.
Experimental results illustrate that the rules initialized by \reflector can better enable rewritten \nl to improve \sql generation accuracy on most \nlsql methods. 
Over the process of multi-round rewriting, the \reflector is able to improve further the performance of subsequent rewriting based on the experience and feedback accumulated from previous rewritten \nl and generation result.

\begin{table*}[t!]
    \centering
    \begin{tabular}{p{2.5cm}p{5cm}p{4cm}p{4cm}}
        \toprule
        \textbf{Flaw Type} & \textbf{Description} & \textbf{Flawed \nl} & \textbf{Details}\\
        \midrule
        \texttt{Missing Info.} & Users may omit information that is essential to generate correct \sql queries, which may cause the model to fail to generate accurate \sql statements. & What model has the most different versions? & \nl does not refer to which table or column to query.\\
        \texttt{Wrong Entity} & Entities (e.g. table names or column names) mentioned by the user in the \nl may be wrong or not present in the database, which will cause the model to generate invalid \sql queries. & What are the locations and names of all \underline{stations} with capacity between 5000 and 10000? & \textbf{stations} do not exist in the \db, the real entity should be \textbf{stadiums}.\\
        \texttt{Ambiguity} & \nl may have multiple meanings or ambiguous words that make it difficult for the model to generate the correct \sql query. & What are the \underline{names} and release years for all the songs of the youngest singer? & There are multiple similar column names (\textbf{Song Name} and \textbf{Name}) in \db.\\
        \texttt{Non-standard Statement} & Users may use non-standard or colloquial expressions, which may cause the model to fail to understand the user's intention correctly. & Show name, country, and age for all singers ordered by age \underline{from the oldest to the youngest}. & The expression of the second half of the sentence should be modified.\\
        \bottomrule
        \end{tabular}
    \caption{The most common flaws summarized by \reflector and related \nl examples.}
    \label{table:discussion_error}
\end{table*}

\begin{table*}[t!]
    \centering
    \begin{tabular}{p{4cm}p{5cm}p{6.5cm}}
        \toprule
        \textbf{Action Type} & \textbf{Description} & \textbf{Rewritten \nl}\\
        \midrule
        \texttt{Complete Information} & Add omitted key information entities to \nl to ensure that the generated \sql query is accurate. & Which \underline{\textbf{car model}} in the \underline{\textbf{model list}} table has the highest number of distinct versions in the \underline{\textbf{cars data}} table?\\
        \texttt{Correct Entities} & Fix incorrect entity in the \nl with reference of \db, ensuring that all mentioned entities are correct and present in the \db. & What are the locations and names of all \underline{\textbf{stadiums}} with capacity between 5000 and 10000?\\
        \texttt{Disambiguation} & Reformulate the ambiguous keywords in \nl in conjunction with the \db, and ensure that each word has a clear meaning. & Show the \underline{\textbf{Song Name}} and \underline{\textbf{Song release year}} of the song by the singer with the lowest Age\\
        \texttt{Normalize Statement} & Convert non-standard or colloquial expressions into standard language that the model can understand. & Show name, country, and age for all singers ordered by age \underline{\textbf{in descending order}}. \\
        \bottomrule
        \end{tabular}
    \caption{The most effective action experience summarized by \reflector and the rewritten \nl for the examples in Table \ref{table:discussion_error}.}
    \label{table:discussion_action}
\end{table*}

\subsection{Discussion}
Based on the aforementioned experiments, several experiences and heuristic insights are as follows:
\begin{itemize}
    \item Table \ref{table:discussion_error} and Table \ref{table:discussion_action} enumerate common flaw and corresponding rewriting actions for flawed \nl summarized by \reflector. 
    During the process of \nlsql, these flaws can lead to misunderstandings in the database, query misinterpretations, or other issues, ultimately resulting in reduced accuracy of \sql generation. The rewriting actions implemented by \rewritermethod improve the performance of associated \nlsql methods by addressing these flaws.
    \item \reflector does not initialize too many experiences, as this could result in duplicate or similar experience descriptions. For example, \texttt{ambiguity} and \texttt{fuzzy statements} actually refer to the same flaw that the given \nl have multiple meanings or are ambiguous.
    \item The running and testing of the \rewritermethod framework and above \nlsql method resulted in a significant consumption of tokens, estimated to be around 45 million tokens. Due to experimental cost constraints, more \nlsql methods cannot be plugged into \rewritermethod. 
\end{itemize}

\section{Related Work}
\paragraph{NL2SQL Translation} The recent advancement of LLM contributed to the progress of \nlsql translation.  Researchers are able to construct \nlsql models through prompt engineering, supervised fine-tuning, and other innovative techniques. The current research focuses on designing and optimizing \nlsql workflow modules, including pre-processing modules (e.g., schema linking~\cite{resdsql}, database content representation~\cite{chess, codes}), translation strategies (e.g design of encoder and decoder~\cite{catsql, ratsql}, intermediate representations~\cite{scprompt, zeronl2sql, natsql, wolfson2020break, eyal2023semantic, gao2022towards, petsql}), and post-processing module (e.g., \sql correction~\cite{dinsql, macsql}, self-consistency~\cite{dailsql, c3sql}), with limited attention given to the original input \nl.
These approaches might ignore potential ambiguities and missing details within the original \nl~\cite{DBLP:conf/sigmod/TangLOLC22, DBLP:conf/sigmod/Luo00CLQ21, DBLP:journals/tvcg/LuoTLTCQ22, DBLP:conf/sigmod/LuoQ00W18, DBLP:journals/corr/abs-2109-03506, DBLP:journals/vldb/QinLTL20}.
In real-world situations, users' understanding of the database can significantly impact the quality of \nl, potentially leading to model errors such as hallucinations or incorrect \sql generation.

In this paper, our goal is to rewrite the flawed \nl, minimize the misrepresentations and ambiguous statements in the \nl, and fix the gap between \nl and \nlsql model, which can enhance the robustness and performance of \sql generation.

\paragraph{Text Rewriting} Text rewriting methodologies aim to assist users in enhancing the quality of their input \nl or adapting the narrative style to suit various contexts. 
In the existing research, text rewriting is extensively employed for privacy protection~\cite{igamberdiev2023dp}, text summary~\cite{bao2023general}, heterogeneous data representation~\cite{wu2023retrieve}, and various other tasks.
In addition, several studies have explored the possibility of text rewriting to improve the quality of \nl and align user's intent~\cite{hwang2023rewriting, shu2024rewritelm}.

Based on the above references, this study aims to apply text rewriting techniques to \nlsql scenarios. The objective is to guide \nlsql models towards generating SQL queries that better align with the user's intent, by enhancing the quality and coherence of \nl of users.

\paragraph{LLM Agents} 
LLM-based agents framework is designed for the simulation of interactive behavior of specific targets or assisting users in solving specified tasks~\cite{Park2023GenerativeAgents, hua2023war, wang2023survey, DBLP:journals/corr/abs-2406-07815}. It usually comprises various modules, such as observation, memory, planning, and response~\cite{zhou2023agents, DBLP:journals/corr/abs-2410-10762}. In NLP-related applications, existing research usually designs different agents in the form of task decomposition and utilizes data stream to make them collaborative~\cite{chess, macsql, li2023tradinggpt}, which can reduce the processing difficulty of sub-tasks and improve the accuracy of generation.

In this research, we propose \rewritermethod, a plug-and-play multi-agent framework to rewrite the user's \nl. It comprises three agents with different divisions of labor, jointly committed to the check-reflection-rewrite workflow. 

\section{Conclusion}
In this paper, we propose \rewritermethod, a novel plug-and-play framework designed to enhance \nlsql systems by automatically rewriting flawed \nl inputs. By leveraging database knowledge such as column values and foreign keys, \rewritermethod reduces translation errors and improves \sql generation accuracy. Extensive experiments on the Spider and BIRD benchmarks demonstrate the effectiveness of \rewritermethod, achieving consistent improvements in execution accuracy by 1.6\% and 2.0\%, respectively. 

\bibliography{ref}

\clearpage

\appendix

\begin{figure*}[!tb]
\section{Appendix}
\subsection{Prompt of Agents}
\end{figure*}

\begin{figure*}[!tb]
    \begin{promptbox}{Prompt of the \checker}
        \begin{verbatim}  
#### Task Description:
Based on the given database schema, Natural Language (NL), SQL query and execution result (only 
showing top three records), determine whether the SQL query is expected to return the correct 
results. You need to follow the steps below for step-by-step reasoning:
1. Syntax Check: Verify if the SQL query adheres to the basic SQL syntax rules.
2. Logical Verification: Extract the database schema information involved in the SQL query.
Based on the given guidelines below, step-by-step determine whether the corresponding SQL
logic matches the expectations of the NL.
3. Execution Analysis: Based on the result set returned from executing the SQL query in the 
real database, verify whether it meets the requirements of the NL.
4. Ambiguity Detection: Check if there is any semantic ambiguity in the NL and whether the 
tables or columns matching the NL in the SQL query are ambiguous. If any ambiguity exists, 
explain the possible ambiguities and directly determine that the SQL query is incorrect.
5. Final Determination: If no issues are found in the above steps, predict the SQL query as 
correct; otherwise, predict the SQL query as incorrect.

### Output Format:
{
  "details": <YOUR THINKING DETAILS>,
  "result": <"TRUE" OR "FALSE">
}

### INPUT:
SCHEMA:  # Fill the database content
{SCHEMA_SLOT} 
NL:   # Fill the NL
{NL_SLOT}
SQL:  # Fill the generated SQL
{SQL_SLOT}
EXECUTION RESULT:  # Fill the execution result
{EXECUTION_RESULT_SLOT}

### OUPUT:
        \end{verbatim} 
    \end{promptbox}
\end{figure*}

\begin{figure*}[!tb]
    \begin{promptbox}{Prompt of the \reflector - Initial Action Experience Generation}
        \begin{verbatim}  
#### Task Description:
Please think deeply about the corresponding modification operation and the corresponding operation
description according to the following problems of the user text, and output in JSON format.

### Output Format:
{
    'Operation Type': 'Description',
    'Operation Type': 'Description',
    ...
}

### The types of problems are:
{ERROR_SPACE}
        \end{verbatim} 
    \end{promptbox}
\end{figure*}

\begin{figure*}[!tb]
    \begin{promptbox}{Prompt of the \reflector - Initial Error Experience Generation}
        \begin{verbatim}  
#### Task Description:
Please list 10 possible problems with user-provided text in the Text-to-SQL task and the 
corresponding explanation of the problem), and output them in JSON format:

### Output Format:
{
    'Problem Type': 'Description',
    'Problem Type': 'Description',
    ...
}
        \end{verbatim} 
    \end{promptbox}
\end{figure*}

\begin{figure*}[!tb]
    \begin{promptbox}{Prompt of the \reflector}
        \begin{verbatim}  
#### Task Description: 
As the AI assistant, your task is to analyze the flaws of the questions (NL)
entered by the user, given the database information and examples. This NL cannot be properly 
translated into a SQL query correctly.
The goal of this task is to provide an analysis and recommendations for a given NL that
can be modified and optimized. To do this well, you need to look for the following details 
in the question:
{FLAW_SLOT}  # Fill the flaw experience

Please analyze whether there are above flaws in the above details in the NL. If so, please select the
available rewriting actions for the NL. The available rewriting actions are as follows:
{ACTION_SLOT} # Fill the action experience

Please clearly mark the keywords in each statement and the corresponding modification suggestions.
### Output Format:
{
  "reflection": <YOUR THINKING DETAILS>
}
### Here are a reflection example:
{
    "reflection": 
        "The flaw is... 
        (describe the specific FLAW with DB), 
        the recommended operation is... , 
        (describes the ACTION with DB); 
        The flaw is... 
        (describe the specific FLAW with DB), 
        the recommended operation is... , 
        (describes the ACTION with DB); 
        ..." 
}

### INPUT:
SCHEMA: # Fill the Database content
{SCHEMA_SLOT}
NL:     # Fill the flawed NL
{NL_SLOT}

### OUPUT:
        \end{verbatim} 
    \end{promptbox}
\end{figure*}

\begin{figure*}[!tb]
    \begin{promptbox}{Prompt of the \rewriter}
        \begin{verbatim}  
#### Task Description:
As the AI assistant, your task is to rewrite the NL entered by the user based on the given 
database information and reflection. 
This NL has some flaws and got bad generation in the downstream models, so you need to make this 
NL as reliable as possible.
The rewritten NL should express more complete and accurate database information requirements 
as far as possible. In order to do this task well, you need to follow these steps to think and 
process step by step:
1. Please review the given reflection and DB information, and first check whether the NL contains 
the corresponding key information and the corresponding flaws. If they exists, please modify, 
supplement or rewrite it in the statement of NL by combining the reflection and DB.
2. Please rewrite the original NL based on the above process. On the premise of providing more 
complete and more accurate database information, the structure of the rewritten NL should be similar 
to the original statement as far as possible. All rewritten statements do not allow delimiters, 
clauses, additional hints or explanations. DONT CONVERT IT INTO QUERY.

{
  "details": <YOUR STEP-BY-STEP THINKING DETAILS>,
  "result": <YOUR FINAL REWRITED NL>
}

### INPUT:
SCHEMA: # Fill the database content
{SCHEMA_SLOT}
NL:     # Fill the flaw NL
{NL_SLOT}
REFLECTION: # Fill the reflection
{REFLECTION_SLOT}

### OUTPUT:
        \end{verbatim} 
    \end{promptbox}
\end{figure*}

\begin{figure*}
\subsection{Description of \nlsql Baselines}
    \begin{itemize}
        \item \textbf{DAIL-SQL}~\cite{dailsql} DAIL-SQL encodes structural knowledge and selects the corresponding few-shot prompt by calculating the similarity order of the skeleton. It also enhances reasoning efficiency by blocking cross-domain specific words in the representation.
        \item \textbf{C3}~\cite{c3sql} implements schema filtering in the schema linking process, and uses prompt calibration and self-consistency to ensure the stability of \sql generation and reduce inference costs
        \item \textbf{DTS-SQL}~\cite{dtssql} is designed to address user privacy data optimization. It comprises two main sub-tasks: schema linking and \sql generation. A two-stage fine-tuning approach is implemented to effectively align the performance of the open-source LLM with that of the proprietary LLM.
        \item \textbf{NatSQL + T5}~\cite{comp} is a semantic-boundary-based technique, which is based on semantic boundaries and involves the use of special tags to identify aligned semantic boundaries between the source problem and the target SQL. 
        \item \textbf{RESDSQL}~\cite{resdsql} comprises a ranking-enhanced encoder and a skeleton-aware decoder. The encoder selectively incorporates the most pertinent schema items, as opposed to the complete set of unordered schema items. Meanwhile, the decoder initially produces the framework and subsequently the specific \sql query, thereby implicitly constraining \sql parsing.
        \item \textbf{GPT-3.5 \& GPT-4}~\cite{openaigpt4} uses chain-of-thought and zero-shot techniques to construct schema linking and generation prompts for the processing and the generation of the \sql.
    \end{itemize}
\end{figure*}

\end{document}